\documentstyle[12pt]{article}

\newcommand{\be}{\begin{equation}}
\newcommand{\ee}{\end{equation}}
\begin{document}
 \begin{flushright}
CERN-TH/96-51\\
 IHES/P/96/12~~~~\\
 UWThPh-14-1996\\
 hep-th/9602115
 \end{flushright}

\begin{center}
{\Large \bf On Finite 4D Quantum Field Theory}\\
{\Large \bf in Non-Commutative Geometry}\\[1pt]
H. Grosse\footnote{Participating in Project No. P8916-PHY
 of the `Fonds zur F\"orderung der wissenschaftlichen Forschung
 in \"Osterreich'.}  \\
{\small Institut for Theoretical Physics, University of Vienna, \\
Boltzmanngasse 5, A-1090 Vienna, Austria \\}
C. Klim\v{c}\'{\i}k\footnote{Partially supported by the grant GA\v CR
 210/96/0310.} \\
{\small Theory Division CERN,
CH-1211 Geneva 23, Switzerland \\}
P. Pre\v{s}najder \\
{\small Department of Theoretical Physics, Comenius University \\
Mlynsk\'{a} dolina, SK-84215 Bratislava, Slovakia\\}
{\bf Abstract} \\
\end{center}

\noindent The truncated 4-dimensional sphere $S^4$ and
 the action of the self-interacting scalar field  on it are constructed.
The path integral quantization is performed while simultaneously keeping
the SO(5)  symmetry and the finite number of degrees of freedom. 
The usual field theory UV-divergences are manifestly absent.

 \noindent CERN-TH/96-51

 \noindent February 1996

\newpage

\section{Introduction}
The basic ideas of  non-commutative geometry were developed in [1, 2],
and in the form of the matrix geometry in [3, 4]. The applications to
physical models were presented in [2, 5], where the non-commutativity
was in some sense minimal: the Minkowski space was not extended
by some standard Kaluza-Klein manifold describing internal degrees of freedom,
but just by two discrete  points. 
The algebra of functions on this manifold remains commutative,
but the complex of the differential forms does not.
This led to a new insight on the
$SU(2)_L \bigotimes U(1)_R$ symmetry of the standard model of electroweak
interactions. The consideration of gravity was included  in [6].  Such models,
of course, do not lead to UV-regularization, since they do not introduce
any modification of the space-time short-distance behaviour.

To achieve the UV-regularization one should introduce a non-commuta-
\newline
tive deformation of the algebra of functions  on a 
 space-time manifold in the Minkowski case, or on
the space manifold in the Euclidean version. One of the simplest locally
Euclidean manifolds is the sphere $S^2$. Its non-commutative (fuzzy) deformation
was described by [7,8] in the framework of the matrix geometry. A more general
construction of some non-commutative homogenous spaces was described in [9]
using coherent-states technique.

The first attempts to construct fields on a truncated sphere were presented in
[8,10] within the matrix formulation. Using a more general approach, the fields
on truncated $S^2$ were investigated in detail in [11--13]. In particular, in
[11] it was the quantum scalar field  on the truncated $S^2$ 
and it was explicitly demonstrated that the UV-regularization automatically
takes place upon the non-commutative deformation of the algebra of functions.

In this article we extend this approach from the 2-dimensional sphere $S^2$
to the 4-dimensional one. Since $S^4$ is not a (co)-adjoint orbit, this
extension has some new nontrivial features. We shall introduce only the
necessary notions of the non-commutative geometry we needed for our approach.

In Sec. 2 we describe briefly the standard (commutative) sphere $S^4$ as
the Hopf fibration $S^7 \to S^4$ and  the scalar self-interacting
field  on it.  Section 3 is devoted to the generalization of the model
to the non-commutative truncated sphere $S^4$ introducing the non-commutative
analogue of the Hopf fibration. Then, using Feynman (path) integrals, we
perform the quantization of the model in question. Last, Sec. 4 contains a
brief discussion and concluding remarks.

\section{Scalar field on the commutative $S^4$}
Here we describe the standard sphere $S^4$ in the form that will be suitable
for the non-commutative generalization. Our basic tools are the real quaternions
\be
\varphi \ =\ \varphi_{(a)} e_a \ \in \ {\bf H}
\ee
with $\varphi_{(a)}$ real and the quaternionic units
\[
e_1 \ =\ \left( \begin{array}{cc}
0 & i \\
i & 0 \end{array} \right) \ ,\
e_2 \ =\ \left( \begin{array}{cc}
 0 & 1 \\
-1 & 0 \end{array} \right) \ ,\
\]
\be
e_3 \ =\ \left( \begin{array}{cc}
i & 0 \\
0 & -i \end{array} \right) \ ,\
e_4 \ =\ \left( \begin{array}{cc}
1 & 0 \\
0 & 1 \end{array} \right) \ ,
\ee
satisfying the relations
\be
e_i e_j \ =\ -\delta_{ij} -\varepsilon_{ijk} e_k \ ,
\ e_4 e_i \ =\ e_i e_4 \ .
\ee
We shall usually write 1 instead of $e_4$. The coefficient
$\varphi_{(0)} = \frac{1}{2} {\rm tr} \phi$  is called the  real part of the
quaternion, and $\varphi_{(i)}$, $i = 1,2,3$, are pure quaternionic components.
The explicit realization (2) of the quaternionic units allows us to identify
the space of quaternions ${\bf }$ with ${\bf C}^2$: any quaternion
we represent by $2 \times 2$ complex matrix
\be
\varphi \ =\ \left( \begin{array}{cc}
\varphi^*_2 & \varphi_1 \\
-\varphi^*_1 & \varphi_2 \end{array} \right) \ .
\ee
The quaternionic conjugation $\varphi \to \varphi^*$ defined by
\[
e_i \to e^*_i = -e_i \ ,\ i = 1,2,3\ ,\ e_4 \to e^*_4 = e_4 \ ,
\]
then corresponds to the Hermitian conjugation of complex matrices. We shall
frequently use both descriptions without an explicit specification. Further,
the quaternionic length $|\varphi |$ is defined by
\be
|\varphi |^2 \ =\ \varphi^* \varphi \ =\ \varphi^2_{(a)} \ =\
{\rm det} \varphi \ .
\ee
If $|\varphi | = 1$, $\varphi$ is called a unit quaternion. The set of unit
quaternions is isomorphic to the group $SU(2)$ (and as a topological space
to $S^3$).

The group $Sp(4)$ we identify with the group of $2 \times 2$ quaternionic
matrices of the form
 \be
A =\ \left( \begin{array}{cc}
\cos \frac{\theta}{2}\ \alpha & \sin \frac{\theta}{2}\ \gamma \beta^* \\
-\sin \frac{\theta}{2}\ \gamma^* \alpha & \cos \frac{\theta}{2}\ \beta^*
\end{array} \right) \ ,
\ee
where $\alpha$, $\beta$, $\gamma$ are unit quaternions, and
$\theta \in [0,\pi ]$ is a real angle.

The Lie algebra $sp(4) = so(5)$ is spanned by 10 anti-Hermitian matrices
$\xi_{AB} = -\xi_{BA}$, $A,B = 1,...,5$, given as
\be
\xi_{a5} \ =\ \left( \begin{array}{cc}
0 & e_a \\
-e^*_a & 0 \end{array} \right) \ =:\ \xi_a \ ,\
\xi_{ab} \ =\ \xi_a \xi_b \ ,
\ee
where $a,b = 1,...,4$, $a \neq b$. The matrices $\xi_{ab}$ span the Lie
algebra $so(4) = so(3) \oplus so(3)$. Supplementing (7) by five  matrices
\be
{\tilde \xi}_a \ =\ \left( \begin{array}{cc}
0 & e_a \\
e^*_a & 0 \end{array} \right) \ ,\
{\tilde \xi}_5 \ =\ \left( \begin{array}{cc}
1 & 0 \\
0 & -1 \end{array} \right) \ ,\
a = 1,...,4\ ,
\ee
we recover the basis of the Lie algebra $su^* (4) = so(5,1)$. It is closely
related to the Clifford algebra $C^{4,0}$ with the basis
$\xi_a$, $a = 1,...,4$:
\be
C^{4,0} =\ \left( \begin{array}{c}
\xi_1 \xi_2 \xi_3 \xi_4 \\
\xi_a \xi_b \xi_c \ ,\ 1\le a<b<c\le 4 \\
\xi_a \xi_b \ ,\ 1\le a<b\le 4 \\
\xi_a \ ,\ 1\le a\le 4 \\
1 \end{array} \right) \
=\ \left( \begin{array}{c}
{\tilde \xi}_5 \\
{\tilde \xi}_a \ ,\ 1\le a\le 4 \\
\xi_{ab} \ ,\ 1\le a<b\le 4 \\
\xi_a \ ,\ 1\le a\le 4 \\
1 \end{array} \right) \ ,\
\ee
where the matrices $\xi_a ,\xi_{ab}$ are anti-Hermitian whereas the matrices
${\tilde \xi}_A$, $ A=1,...,5$, are Hermitian and transform as an $SO(5)$
vector.

The matrices $A \in Sp(4)$ act in a natural way in the space ${\bf H}^2$:
\be
z \ =\ \left( \begin{array}{c}
\varphi \\
\chi \end{array} \right) \ \in \ {\bf H}^2 \ \to
Az \ =\ \left( \begin{array}{c}
{\tilde a}\varphi + {\tilde b}\chi \\
{\tilde c}\varphi + {\tilde d}\chi \end{array} \right) \ \in \ {\bf H}^2 \ .
\ee
The sphere $S^7$, given by the equation
\be
z^+ z \ =\ |\varphi |^2 \ +\ |\chi |^2 \ =\ 1\ ,
\ee
is transitively invariant under this action. Introducing the equivalence
relation
\be
z\ \sim \ z' \ =\ z \alpha \ ,\ \alpha \ -\ {\rm unit}\ {\rm quaternion}\ ,
\ee
we recover the sphere $S^4$ as the Hopf fibration $S^7 \to S^4$.
To any equivalence class (13) we assign the $SO(5)$ vector given by the
Cartesian coordinates in ${\bf R}^5$:
\be
x_A \ =\ \frac{1}{2} {\rm tr} (z^+ {\tilde \xi}_A z) \ =\
\frac{1}{2} {\rm tr} ({z'}^+ {\tilde \xi}_A z' ) \ .
\ee
These are just the Cartesian coordinates of the sphere $S^4$ embedded into
${\bf R}^5$ (similar objects were used in [8] within a relativistic
context).

As ${\cal A}_{\infty}$ we denote the commutative algebra of analytic
functions (polynomials) in the variables
$x_A$, $A=1,...,5$:
\be
\Phi (x)\ =\ \sum A_M x^M \ ,\ A_M-{\rm complex} \ ,
\ee
with the usual point-wise multiplication. Here we used the multi-index
notation: $M=(M_1 ,...,M_5 )$, $x^M = x_1^{M_1} ... x_5^{M_5}$.  In
${\cal A}_{\infty}$ we introduce  the scalar product
\be
(\Phi_1 ,\Phi_2 )_{\infty} \ =\ I_{\infty} [\Phi^*_1 \Phi_2 ]\ ,
\ee
where $I_{\infty} [...]$ denotes the usual $SO(5)$-invariant integral on $S^4$:
\be
I_{\infty} [...] \ =\ \frac{3}{4\pi^2} \int d^5 x\ \delta (x^2_A -1)\ [...]
\ ,
\ee
where the normalization guarantees that $I_{\infty} [1] = 1$.

The $Sp(4)$ action (10) in the algebra ${\cal A}_{\infty}$ generates
${\bf R}^5$ rotations, leaving the quantity $x_A^2 = 1$ invariant. The
generators of this action (anti-Hermitian with respect to the scalar product
given above) are given as
\be
{\hat J}_{AB} \Phi \ =\ \frac{1}{2} (\psi^*_{\alpha} \xi^{\alpha \beta}_{AB}
\partial_{\psi^*_{\beta}} \ +\ \psi_{\beta} \xi^{\alpha \beta}_{AB}
\partial_{\psi_{\alpha}} )\Phi \ .
\ee
Here $\xi^{\alpha \beta}_{AB}$ are elements of the $4 \times 4$ complex
matrix assigned to the $2 \times 2$ quaternionic matrix $\xi_{AB}$, and
$\psi_{\alpha}$, $\psi^*_{\alpha}$, $\alpha =1,...,4$, are complex variables
identified with the elements of complex matrices assigned to the quaternions
$\varphi$ and $\chi$ in the
following way:
\[
\psi_1 \ =\ \varphi_1 \ ,\ \psi_2 \ =\ \varphi_2 \ ,\
\psi_3 \ =\ \chi_1 \ ,\ \psi_4 \ =\ \chi_2 \ ,
\]
\be
\psi^*_1 \ =\ \varphi^*_1 \ ,\ \psi^*_2 \ =\ \varphi^*_2 \ ,\
\psi^*_3 \ =\ \chi^*_1 \ ,\ \psi^*_4 \ =\ \chi^*_2 \ .
\ee

It follows from (17) that the quantities $\psi_{\alpha}$, $\psi^*_{\alpha}$,
$\alpha =1,...,4$, transform as $S^4$ spinors
\be
{\hat J}_{AB} \psi_{\alpha} \ =\
\frac{1}{2} \xi^{\alpha \beta}_{AB} \psi_{\beta} \ , \
{\hat J}_{AB} \psi^*_{\beta} \ =\
\frac{1}{2} \xi^{\alpha \beta}_{AB} \psi^*_{\alpha} \ .
\ee
Consequently, the quantities $x_A$, $A=1,...,5$ given as
\be
x_A \ =\ \psi^+ {\tilde \xi}_A \psi \ =\
\psi^*_{\alpha} {\tilde \xi}^{\alpha \beta}_A \psi_{\beta} \ ,
\ee
where ${\tilde \xi}^{\alpha \beta}_A$ are elements of the complex
matrix assigned to ${\tilde \xi}_A$, transforms as a vector in ${\bf R}^5$.
Moreover, the function $C(x) = x^2_A$ satisfies
\be
{\hat J}_{AB} C(x) \ =\ 0\ ,\ A,B =1,...,5 \ ,
\ee
i.e. $C(x)$ is an invariant function as expected.

The $Sp(4)$ action (17) in the algebra ${\cal A}_{\infty}$ is reducible and
we have the following expansion:
\be
{\cal A}_{\infty} \ =\ \bigoplus_{p=0}^{\infty} \ {\cal A}^{p}_{\infty} \ ,
\ee
where ${\cal A}^{p}_{\infty}$ is the carrier space of the irreducible
representation of the $Sp(4)$ group spanned by the harmonic polynomials
$\Psi^p_{\mu}$ of degree $p$ in the variables $x_A$, $A=1,...,5$. The
polynomials $\Psi^p_{\mu}$ are orthonormal with respect to the scalar
product (15). The dimension of the space ${\cal A}^{p}_{\infty}$ is
\[
d_p \ =\ {\rm dim} {\cal A}^{p}_{\infty} \ =\
\frac{1}{6} (p+1)(p+2)(2p+3)\ ,
\]
which means that any field $\Phi \in {\cal A}_{\infty}$
can be expanded as
\be
\Phi (x) \ =\
\sum_{p=0}^{\infty} \sum_{\mu =0}^{d_p} a^p_{\mu} \Psi^p_{\mu} \ .
\ee
The field action corresponding to the real scalar field $\Phi$ is given as
\be
S[\Phi ]\ =\
I_{\infty} \Big[\frac{1}{2}({\hat J}_{AB} \Phi )^2 \ +\ V(\Phi )\Big]\ ,
\ee
where $V(.)$ is a polynomial bounded from below.

The quantum mean value of some polynomial field functional $F[\Phi ]$
is defined as the functional integral over fields from
$\Phi \in {\cal A}^R_{\infty}$ by
\be
\langle F[\Phi ] \rangle =
\frac{\int D\Phi e^{-S[\Phi ]} F[\Phi ]}{\int D\Phi e^{-S[\Phi ]}} \ ,
\ee
where $D\Phi = \prod_x d\Phi (x) = \prod_{p,\mu} da^p_{\mu}$ (eventually,
with the reality conditions for $a^p_{\mu}$ included).

Since here $p = 0,1,...,\infty$, the formula for the measure is only formal.
We shall not discuss the complicated (and not completely solved) problems 
related
to its rigorous definition. As we shall see below, such problems do not appear
in the framework of the non-commutative version of the model.

\section{Scalar field on the non-commutative  $S^4$}
In this section we shall use various unitary irreducible representations of
the group $Sp(4)$. Any such representation is characterized by its signature
$(p,k)$ with integer $p \ge k \ge 0$ and can be expressed as the Young product
\be
(p,k)\ =\ \pi_1^{p-k} \ \pi_2^k \ ,
\ee
of $Sp(4)$ fundamental representations: $\pi_1 = (1,0)$  4-dimensional
quaternionic and $\pi_2 = (1,1)$ 5-dimensional real  (see e.g. [14]). The
dimension of the representation $(p,k)$ is
\be
d_{pk} \ =\ \frac{1}{6} (p+2)(k+1)(p-k+1)(p+k+3) \ .
\ee

In the non-commutative (fuzzy) case we replace the commuting parameters (18)
by the non-commutative ones. Namely, we shall express the parameters
$\psi_{\alpha}$, $\psi^*_{\alpha}$, $\alpha = 1,...,4$ in terms of
annihilation and creation operators as
\be
\psi_{\alpha} \ =\ A_{\alpha} R^{-1/2} \ ,\
\psi^*_{\alpha} \ =\ R^{-1/2} A^*_{\alpha} \ ,
\ee
where
\be
R \ =\ A^*_{\alpha} A_{\alpha} \ ,
\ee
so that the condition $\psi^*_{\alpha} \psi_{\alpha} = 1$ is satisfied (the
operators $\psi_{\alpha}$ are well defined everywhere, except in the 
vacuum; we complete the
definition by postulating that they annihilate the vacuum). The operators
$A_{\alpha}$ and $A^*_{\alpha}$ (* denotes the Hermitian conjugation) act in
the Fock space ${\cal F}$ spanned by the orthonormal vectors $|n \rangle =
|n_1 ,...,n_4 \rangle$ labelled by the occupation numbers $n_{\alpha}$,
$\alpha = 1,...,4$. They satisfy in ${\cal F}$ the commutation relations
\be
[A_{\alpha} , A_{\beta} ]\ =\ [A^*_{\alpha} , A^*_{\beta} ]\ =\ 0\ ,
[A_{\alpha} , A^*_{\beta} ]\ =\ \delta_{\alpha \beta} \ .
\ee

The operators
\be
J_{AB} \ =\
\frac{1}{2} A^*_{\alpha} \xi^{\alpha \beta}_{AB} A_{\beta} \ ,\
A,B = 1,...,5 \ .
\ee
satisfy in the Fock space $\cal F$ the $sp(4) = so(5)$ Lie algebra commutation
relations. The subspace ${\cal F}_N$ with the fixed total occupation number
\be
{\cal F}_N \ =\ \{ |n \rangle \ ,\ |n| \ =\ N\ \}\ ,\ N\ =\ 0,1,2,...
\ee
has the dimension
\be
d_{N0} \ =\
\left( \begin{array}{c}
N+3 \\
3 \end{array} \right) \ .
\ee
and is the carrier space of the $Sp(4)$ unitary irreducible representation $(N,0)$.

As the ${\cal A}_N$ we denote the non-commutative algebra of operators
${\cal F}_N \to{\cal F}_N$, which can be expressed as polynomials
\be
\Phi (x)\ =\ \sum A_M x^M \ ,\ A_M {\rm complex} \ ,
\ee
in operators
\be
x_A \ =\ 
\psi^*_{\alpha} {\tilde \xi}^{\alpha \beta}_A \psi_{\beta} \ =\
\psi^+ {\tilde \xi}_A \psi \ ,\ A = 1,...,5 \ 
\ee
restricted to the space ${\cal F}_N$. The operators $x_A$, $A = 1,...,5$,
form a vector in ${\bf R}^5$.

In ${\cal A}_N$ we introduce the scalar product
\be
(\Phi_1 ,\Phi_2 )_N \ =\ I_N [\Phi^*_1 \Phi_2 ]\ ,
\ee
where $I_N [...]$ is the analogue of the  $SO(5)$-invariant integral on $S^4$:
\be
I_N [...] \ =\ \frac{1}{d_{N0}} {\rm Tr}_N [...] \ .
\ee
Here ${\rm Tr}_N [...]$ denotes the trace in the algebra ${\cal A}_N$, and
the normalization guarantees that $I_N [1] = 1$.

As a non-commutative analogue of (18) we have a commutator action of the
$sp(4)$ algebra in ${\cal A}_N$:
\be
{\hat J}_{AB} \Phi (x) \ =\ [ J_{AB} ,\Phi (x)]\ ,
\ee
with $J_{AB}$ defined in (31). This is a reducible representation with the
following decomposition to $Sp(4)$ irreducible components:
\be
(N,0) \otimes (N,0) \ =\ \bigoplus_{p=0}^N \bigoplus_{k=0}^p \ (p+k,p-k) \ .
\ee
This decomposition induces the decomposition of the algebra ${\cal A}_N$:
\be
{\cal A}_N \ =\ \bigoplus_{p=0}^N \bigoplus_{k=0}^p \
{\cal A}^{p+k,p-k}_N \ ,
\ee
where ${\cal A}^{p' k'}_N$ is the carrier space of the $Sp(4)$ representation
$(p' ,k' )$. This means that any $\Phi \in {\cal A}_N$ can be expanded as
\be
\Phi (x) \ =\ \sum_{p=0}^N \sum^p_{k=0} \sum_{\mu =1}^{d'_{pk}}
a^{p+k,p-k}_{\mu} \Psi^{p+k,p-k}_{\mu} \ ,
\ee
where $d'_{pk} = d_{p+k,p-k}$ and $\Psi^{p' k'}_{\mu}$,
$\mu = 1,...,d_{p' k'}$, span the space ${\cal A}^{p' k'}_N$.

In the commutative case, the decomposition (22) of the algebra
${\cal A}_{\infty}$ contains only representations $(p,p) = \pi_2^p$
corresponding to terms with $k=0$ in the decomposition (40).

{\it Note}: We would like to stress that it is not essential that the
generators $x_A$, $A=1,...,5$, given in (35) do not close to some Lie algebra
(they close to a Lie algebra only after supplementing them by the operators
(31)). The following point is important, however: the decomposition (40) of the basic algebra
${\cal A}_N$ under the 
symmetry transformation in question (this aspect was less
transparent for the truncated sphere $S^2$, since in this case the generators
closed to a Lie algebra, see [11]). The detailed information contained in Eq.
(40) is necessary for realistic numerical or symbolical calculations.

We identify the space of the  configurations of a real scalar field 
with the subspace
\be
{\cal A}^R_N \ =\ \bigoplus_{p=0}^N \ {\cal A}^{pp}_N \ ,
\ee
of symmetric polynomials in $x_A$, $A = 1,...,5$, with real coefficients.

Such fields can be expanded as
\be
\Phi (x) \ =\ \sum_{p=0}^N \sum_{\mu =1}^{d'_{pk}}
a^p_{\mu } \Psi^{pp}_{\mu} \ ,
\ee
where the coefficient $a^p_{\mu}$ are real provided that $\Psi^{pp}_{\mu}$
are chosen to be Hermitian (if this is not the case the coefficients
$a^p_{\mu}$ satisfy some relations that guarantee that the field in question
is a Hermitian operator in ${\cal F}_N$). This guarantees that in the
commutative limit $N \to \infty$ we recover from (43) only  fields that have
the proper form (23).

In the non-commutative case the field action corresponding to the real scalar
field $\Phi$ is given as
\be
S[\Phi ]\ =\
I_N \Big[\frac{1}{2} ({\hat J}_{AB} \Phi )^2 \ +\ V(\Phi )\Big] \ ,
\ee
where $V(.)$ is a polynomial bounded from below. Obviously, this action
has the following basic properties:

1) it has the full $SO(5)$ symmetry corresponding to $S^4$ rotations, and

2) it describes a model with a finite number of modes since, in fact, it
corresponds to a particular matrix model.

The quantum mean value of some polynomial field functional $F[\Phi ]$
is defined as the functional integral
\be
\langle F[\Phi ] \rangle =
\frac{\int D\Phi e^{-S[\Phi ]} F[\Phi ]}{\int D\Phi e^{-S[\Phi ]}} \ .
\ee
However, here $D\Phi = \prod_{p,\mu} da^p_{\mu}$ (eventually with the
reality conditions included) is the usual Lebesgue measure, since now the
product is finite ($p = 0,1,...,N$, and $\mu = 1,...,d'_{pp}$). The
quantum mean values are well defined for any polynomial functional
$F[\Phi ]$.

\section{Concluding Remarks}
We have  demonstrated  that the interacting  scalar field on the
noncommutative sphere $S^4$ represents a quantum system that  has the
following properties:

1) The model has the full $SO(5)$ space symmetry  under
the rotations of the sphere $S^4$. This is exactly the same
symmetry as the interacting scalar field on the standard sphere $S^4$ 
possesses.

2) The field has only a finite number of modes. Then the number of degrees
of freedom is finite, which leads to the non-perturbative
UV-regularization, i.e. all quantum mean values of polynomial field
functionals are well defined and finite.

In our approach the UV cut-off in the number of modes is supplemented with a
highly non-trivial vertex modification due to non-trivial products of fields.
Our UV-regularization is non-perturbative and is completely determined by the
algebra ${\cal A}_N$. It is originated by the short-distance structure of
the space, and does not depend on the field action of the model in question.

Moreover, it can be shown that the Schwinger functions
\be
S_n (F) = \langle F_n [\Phi ] \rangle \ ,
\ee
where $F_n[\Phi] = \sum \alpha^{p_1 ...p_n}_{\mu_1 ...\mu_n}
(\Psi^{p_1}_{\mu_1},\Phi )_N\dots (\Psi^{p_n}_{\mu_n},\Phi )_N$ satisfy the
Osterwalder-Schrader axioms:

(OS1) {\it Hermiticity}
\[
S^*_n (F) = S_n (\Theta F) \ ,
\]
where $\Theta F$ is the involution defined by
$\Theta F_n [\Phi] = (F_n [\Phi])^*$.

(OS2) {\it Covariance}
\[
S_n (F) = S_n ({\cal R}F) \ ,
\]
where ${\cal R}F$ is a mapping of functionals induced by $SO(5)$ rotations.

(OS3) {\it Reflection positivity}
\[
\sum_{n,m\in {\cal I}} S_{n+m} (\Theta F_n \otimes F_m ) \geq 0 \ .
\]

(OS4) {\it Symmetry}
\[
S_n (F) = S_n (\pi F) \ ,
\]
where $\pi F$ is a functional obtained from $F$ by arbitrary permutation
of indices of $\alpha^{p_1 ...p_n}_{\mu_1 ...\mu_n}$.

{\it Note}: We do not include the last Osterwalder-Schrader axiom, the
cluster property, since the compact manifold requires a special treatment
(however, it can be recovered in the limit where the radius of the sphere
grows to infinity, but these considerations go beyond the presented scheme).
Qualitatively, the properties of the Schwinger functions are the same
as those valid for the truncated sphere $S^2$, see [11]. We would like to
stress that the properties of standard Schwinger functions not included
above (e.g. support, or singularity and growth, specification) are essential
again in the commutative limit $N \to \infty$.

The usual divergences will appear only in the commutative limit
$N \to \infty$. It would be very interesting to isolate the large-$N$
behaviour non-perturbatively. By this we mean the Wilson-like approach
in which the renormalization group flow in the space of Lagrangians 
is studied. In this context a connection may be found with similar
recent works [15].

Combining the results of this paper with those of [11--13] we obtain a set of
UV-regularized Euclidean quantum field models on $S^2$ and $S^4$:

a) the scalar field on the truncated $S^2$, which is super-renormalizable,

b) the Neveu-Schwarz model on the truncated $S^2$, which is renormalizable,

c) the scalar field on the truncated $S^4$ with $\Phi^4$ interaction which
is renormalizable too.

Analogous models formulated on standard Euclidean planes (${\bf R}^2$ or
${\bf R}^4$ instead of spheres) served as important examples in  the proof
of the existence of quantum fields in continuum Euclidean spaces in the
framework of the Wilson approach (see [16, 17] for the super-renormalizable
case, and [18, 19] for the renormalizable one).

We have an alternative approach: the regularization procedure is
non-perturbative and preserves all space symmetries of the models in
question. The UV-regularization in our scheme can be interpreted as a
direct consequence of the short-distance structure induced by the
non-commutative geometry of the underlying space. This can lead to a
better understanding of the origin and properties of divergences in 
quantum field theory. 
\vskip1pt

{\bf Acknowledgements}~
We are grateful to  A. Connes, V. \v{C}ern\'{y}, M. Fecko, J. Fr\"{o}hlich,
K. Gaw\c edzki, E. Hawkins, B. Jur\v{c}o, J. Madore, R. Stora and 
Ch. Schweigert  for
useful discussions.  P.P. thanks the I.H.E.S. at Bures-sur-Yvette, where  part
 of his research has been done, for hospitality.

\end{document}